\documentclass[12pt]{article}
\usepackage{a4,epsfig,amsmath,amssymb,graphicx,multirow,oldgerm,euscript,mathrsfs}

\begin{document}

\begin{titlepage}
\begin{flushright}
 ${\rm ECT}^{*}$-05-15\\
\end{flushright}
\renewcommand{\thefootnote}{\fnsymbol{footnote}}
\vspace{1.0em}

\begin{center}
{\bf \LARGE{Direct $\boldsymbol{CP}$ violation in $\boldsymbol{B \to \rho^0(\omega) PS}$}}
\end{center}
\vspace{1.0em}
\begin{center}
\begin{large}
O. Leitner$^{1}$\footnote{leitner@ect.it},
Xin-Heng Guo$^{2}$\footnote{xhguo@bnu.edu.cn}, 
A.W. Thomas$^{3}$\footnote{awthomas@jlab.org}
 \\
\end{large}
\vspace{2.3em}
$^1$ ${\rm{ECT}}^{*}$, Strada delle Tabarelle, 286, 
38050 Villazzano (Trento), Italy \\
$I.N.F.N.$, Gruppo Collegato di Trento, Trento, Italy\\
\vspace{0.2cm}
$^{2}$ 
Key Laboratory of Radiation Beam Technology, Beijing, China \\
and \\
Material Modification of National Ministry of Education, Beijing, China \\
and \\
Institute of
         Low Energy Nuclear Physics, Beijing Normal University, China\\
\vspace{0.2cm}
$^{3}$ 
Thomas Jefferson National Accelerator Facility
12000 Jefferson Ave., \\
Newport News VA 23606 USA\\
\end{center}
\vspace{5.0em}
\begin{abstract}
\vspace{1.0em}
We calculate the direct $CP$ violating asymmetry parameter, $a_{CP}$, for $B
\to \pi^{+} \pi^{-} \pi$ and $B \to \pi^{+} \pi^{-} K$ decays, in the case where
$\rho^0-\omega$ mixing effects are taken into account. We find that the direct
$CP$ asymmetry for $B^{-} \to \pi^{+} \pi^{-} \pi^{-}$,
$\bar{B}^{0} \to \pi^{+} \pi^{-}\pi^{0}$, $B^{-} \to \pi^{+} \pi^{-} K^{-}$
and $\bar{B}^{0} \to \pi^{+} \pi^{-}\bar{K}^{0}$, reaches its maximum when the
invariant mass $\pi^{+} \pi^{-}$ is in the vicinity of the $\omega$ meson mass.
The inclusion of  $\rho^0-\omega$ mixing provides an opportunity to erase, without
ambiguity, the phase uncertainty mod$(\pi)$ in the determination of the CKM angles
$\alpha$ in case of $b\to u$ and $\gamma$ in case of $b \to s$.
\end{abstract}
%
\end{titlepage}
\newpage
\section{Direct $\boldsymbol{CP}$ violation}
Direct $CP$ violating asymmetries in $B$ decays occur through the interference of,
at least, two amplitudes with different weak phase, $\phi$, and strong phase, $\delta$.
The extraction of the weak phase $\phi$ (which is determined by a combination
of CKM matrix elements) is made through the measurement of a $CP$ violating asymmetry. However,
one must know the strong phase $\delta$ which is not still well determined in any theoretical
framework. In this regard, the isospin symmetry violating mixing between $\rho^{0}$ and $\omega$ can
be extremely important, since it can lead to a large $CP$ violation in $B$ decays such as
$B \to \rho^0(\omega) Y \to \pi^+ \pi^- Y$  ($Y$ represents a meson) because the strong
phase passes through $90^{o}$ at the $\omega$ resonance. In any phenomenological treatment
of the weak decays of hadrons, the starting point is the weak effective Hamiltonian at low
energy. It is obtained by integrating out the heavy fields from the Standard Model Lagrangian.
The Operator Product Expansion   is used  to separate the calculation of the amplitude,
$A(B \rightarrow F)\propto C_{i}(\mu) \langle F | O_{i} | B \rangle (\mu)$,  into two
distinct physical regimes. One is called {\it hard}, represented by $C_{i}(\mu)$ and
calculated  by a perturbative approach. The other  is called {\it soft}, described by
$O_{i}(\mu)$ and  derived by using a non-perturbative approach. The operators, $O_{i}(\mu)$,
can be understood as local operators which govern effectively a given decay, reproducing
the weak interaction of quarks in a point-like approximation. The Wilson coefficients,
$C_{i}(\mu)$,  represent the physical contributions from scales higher than $\mu (=m_b)$ 
and they can be calculated in perturbation theory because of the property of  asymptotic
freedom of QCD.

Factorization in  $B$ decays involves three fundamental scales: the weak
interaction scale, $M_{W}$,  the $b$ quark mass scale, $m_{b}$, and the strong
interaction scale, $\Lambda_{QCD}$. The QCD factorization (QCDF) approach, based
on the concept of color transparency  as well as on a soft collinear  factorization where the
particle energies are bigger than the scale $\Lambda_{QCD}$,  allows us to write down the
matrix elements  $\langle F | O_{i} | B \rangle(\mu)$ at the  leading order in
$\Lambda_{QCD}/m_{b}$ and $\alpha_s$. The hadronic decay amplitude involves both soft and hard
contributions. At leading order, all the non-perturbative effects are assumed to be contained
in the semi-leptonic form factors and the light cone distribution amplitudes. Then, non-factorizable
interactions are dominated by hard gluon exchanges  and can be  calculated  perturbatively, in
order to correct the naive factorization (NF) approximation. It has been also shown that the weak
annihilation contributions cannot  be neglected in $B$ meson decays even though they are
power suppressed in the heavy-quark limit ($\Lambda_{QCD}/m_{b}$). Their contributions are 
approximated in terms of convolutions of hard scattering kernels with light cone expansions
for the final state mesons. Finally, the perturbative calculation of the hard scattering spectator 
and annihilation contributions is regulated by a physical scale of order $\Lambda_{QCD}$.

The direct  $CP$  violating asymmetry parameter, $a_{CP}$, is found to be small for most of 
the non-leptonic  $B$ decays when either the naive or QCD factorization framework
is applied. However,  in the case of $B$ decay channels involving the $\rho^0$ meson, it appears
that  the asymmetry may be large in the vicinity of $\omega$ meson mass. We stress that   
$\rho^0-\omega$ mixing has the dual advantages that the strong phase difference
is large (passing rapidly through $90^{o}$ at the $\omega$ resonance) and well known.
In the vector meson dominance model, the photon propagator is dressed by coupling
to  the vector mesons $\rho^0$ and $\omega$. In this regard, the $\rho^0-\omega$ mixing
mechanism has been developed. Knowing the ratio, $r$, between the tree and penguin amplitudes,
and the strong phase, $\delta$, as well as the weak phase, $\phi$, from the CKM matrix, 
it is  possible to calculate the $CP$ violating asymmetry, $a_{CP}$, including
the $\rho^{0}-\omega$ mixing mechanism. More detail for all the results presented
here can be found in Ref.~\cite{Leitner:2004ij}.
\section{Isospin symmetry violation and direct $\boldsymbol{CP}$ violation in $\boldsymbol{B}$ decays}
\begin{figure}[hpb]
\includegraphics*[width=0.495\columnwidth]{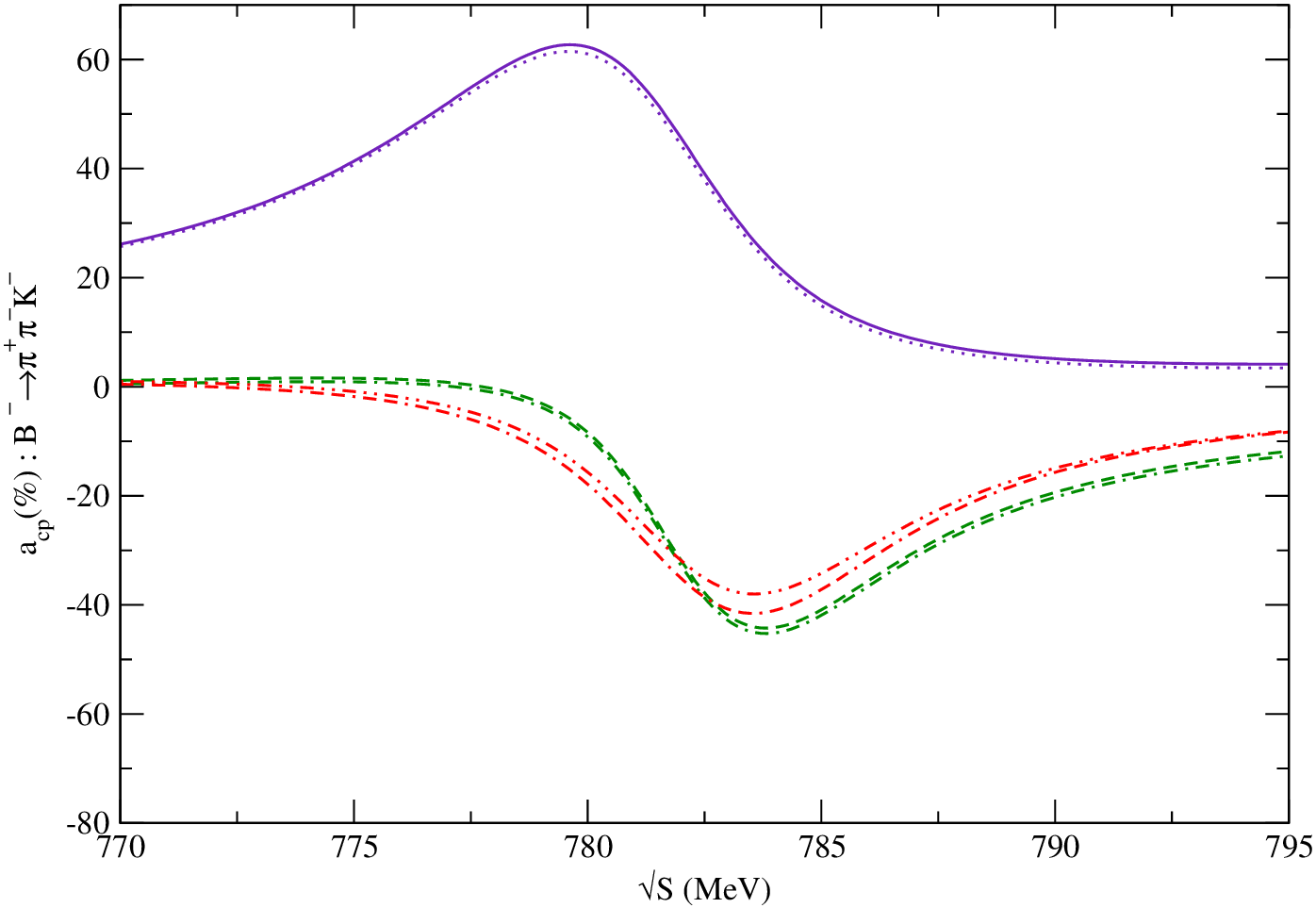}
\includegraphics*[width=0.495\columnwidth]{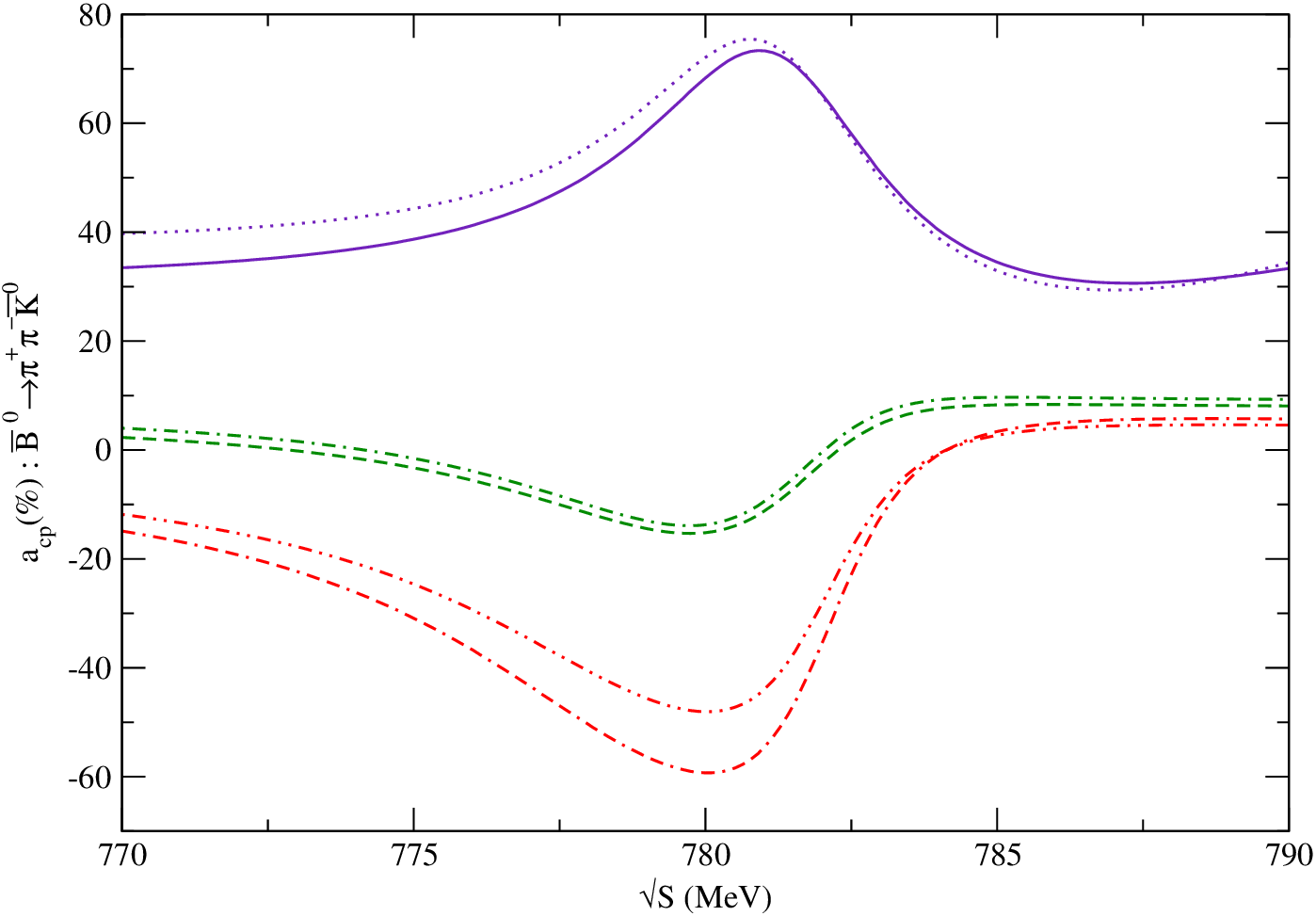}
\includegraphics*[width=0.495\columnwidth]{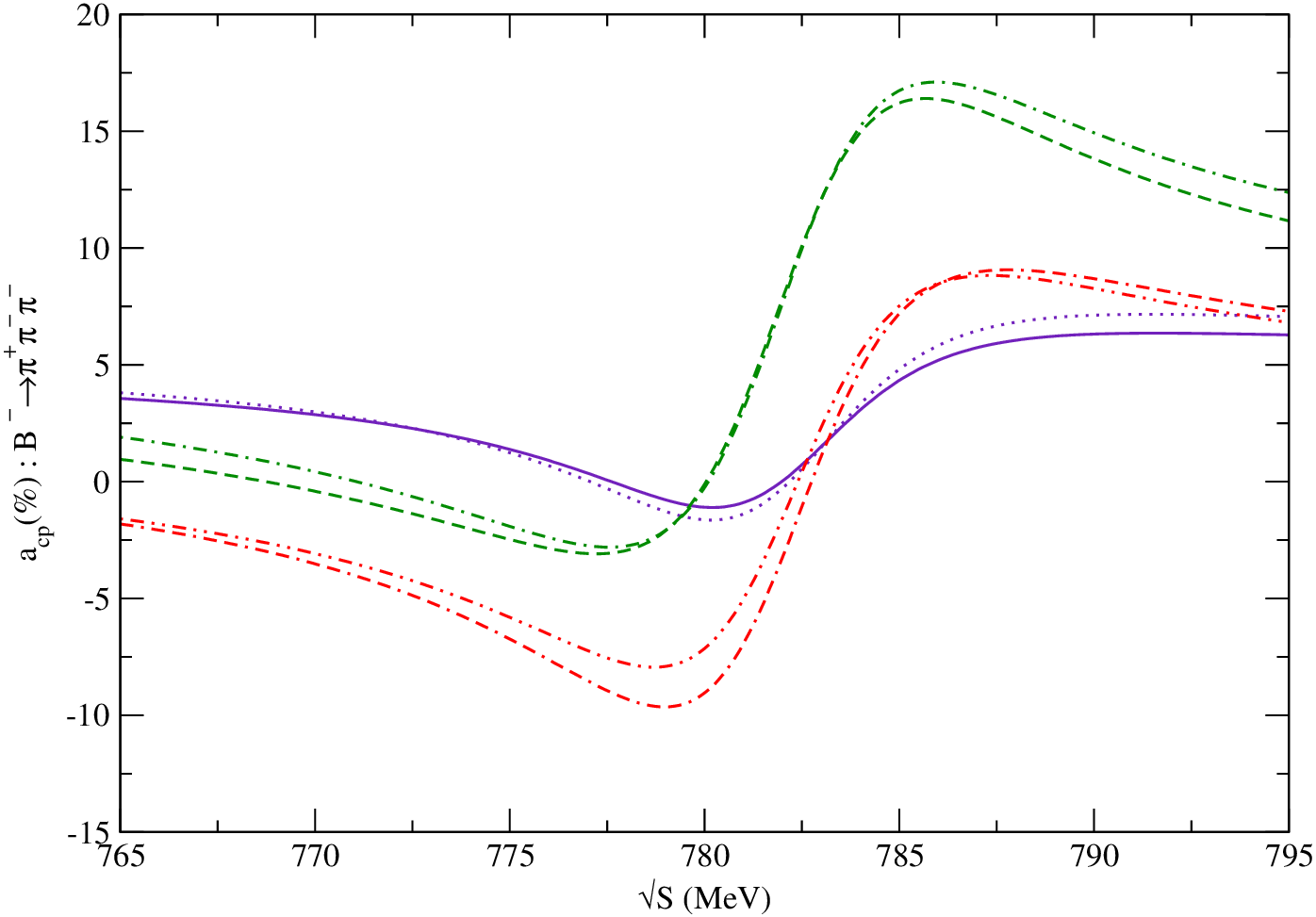}
\includegraphics*[width=0.495\columnwidth]{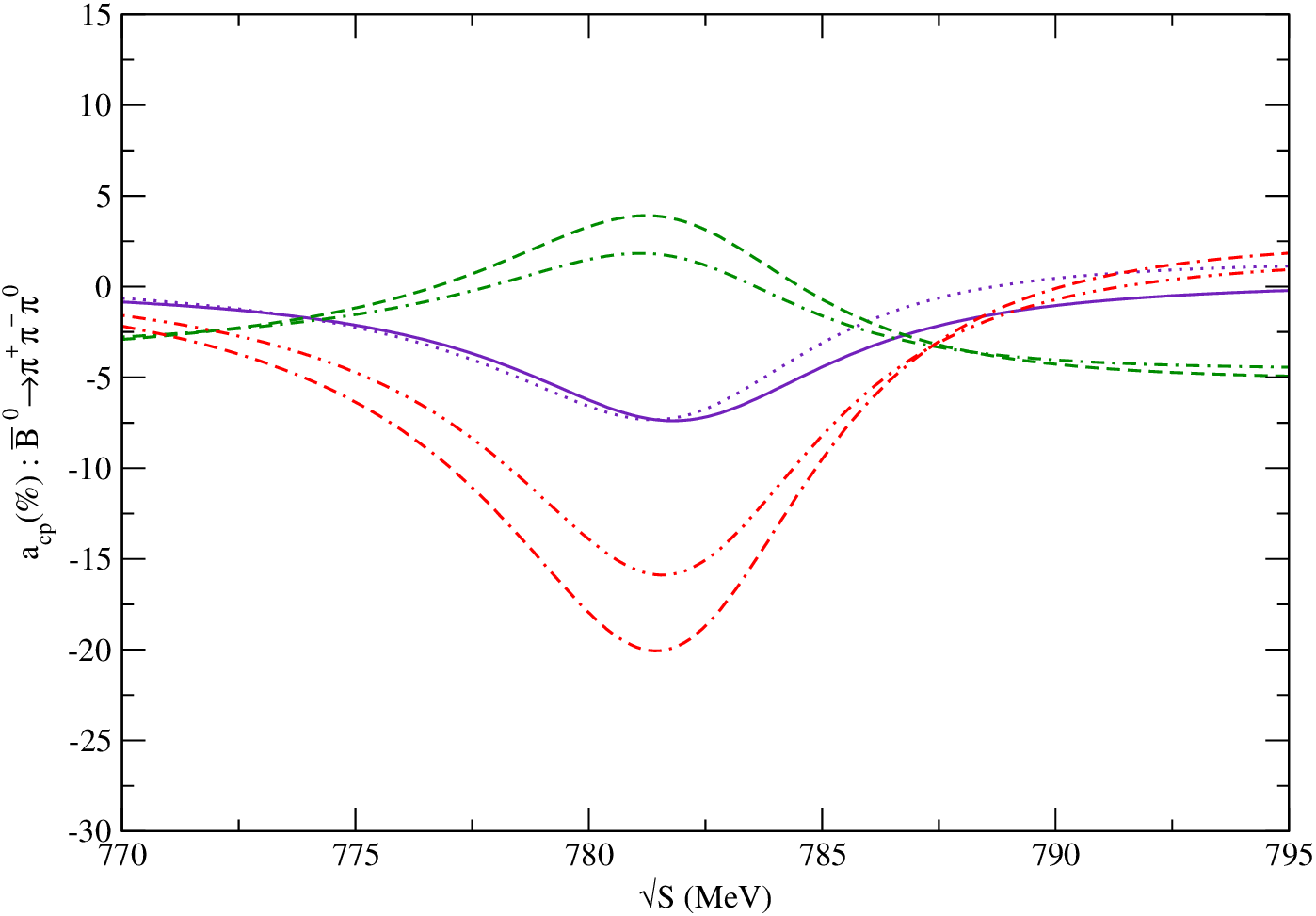}
\caption{First row, $CP$ violating asymmetry, $a_{CP}$, for $B^- \to \pi^+ \pi^- K^-,
\bar{B}^0 \to \pi^+ \pi^- \bar{K}^0$  for max CKM matrix elements. Solid line (dotted line)
for QCDF, dot-dot-dashed line (dot-dash-dashed line) for NF, dot-dashed line (dashed line)
for QCDF with default values and for $F^{B \to K}=0.35(0.42)$. Second row, $CP$ violating
asymmetry, $a_{CP}$, for $B^- \to \pi^+ \pi^- \pi^-, \bar{B}^0 \to \pi^+ \pi^- \pi^0$, for
max CKM matrix elements. Same notation for lines as in first row with $F^{B \to \pi}=0.27(0.35)$.
 All the figures are given as a function of $\sqrt{S}$.}
\label{fig14}
\end{figure}

%
In Fig.~\ref{fig14},  we  show the $CP$ violating asymmetry for $B^{-} \to \rho^{0}(\omega)
\pi^{-}\to \pi^{+} \pi^{-} \pi^{-}$ and  $\bar{B}^{0} \to \rho^{0}(\omega) \pi^{0} \to 
\pi^{+} \pi^{-} \pi^{0}$ respectively, as a function of the energy, $\sqrt{S}$, of the two
pions coming from $\rho^0$ decay, the form factor, $F_{1}^{B \to \pi}$, and the CKM matrix 
element parameters $\rho$ and $\eta$. For comparison, on the same plot we show the $CP$
violating asymmetries, $a_{CP}$, when NF is applied as well as QCDF where default values for the
phases, $\varphi_{H,A}^{M_i}$, and parameters, $\varrho_{H,A}^{M_i}$ are used. In the latter
case, we take $\varphi_{H,A}^{M_i}=0$ and  $\varrho_{H,A}^{M_i}=1$ for all the particles.
Focusing first  on Fig.~\ref{fig14}, where the asymmetry for  $B^{-} \to \rho^{0}(\omega)
\pi^{-}\to \pi^{+} \pi^{-} \pi^{-}$ is plotted, we observe that the $CP$ violating asymmetry 
parameter, $a_{CP}$, can be large  outside the region where the invariant mass of the $\pi^{+}
\pi^{-}$ pair is in   the vicinity of the $\omega$ resonance. This is the first consequence of  
QCD factorization, since within this framework, the strong phase can be  generated not only by
the $\rho^0-\omega$ mechanism but also  by the  Wilson coefficients. Because of the strong 
phase that is either at the order of $\alpha_{s}$ or power suppressed by $\Lambda_{QCD}/m_{b}$,
the $CP$ violating asymmetry, $a_{CP}$, may be small but a large asymmetry cannot be excluded. 
At the $\omega$ resonance, the asymmetry parameter, $a_{CP}$, for $B^- \to \pi^+ \pi^- \pi^-$, is
around $0\%$ in our case. In comparison, the asymmetry  parameter,  $a_{CP}$,  (still at the 
$\omega$ resonance) obtained by applying the naive factorization gives  $-10\%$ whereas it
gives $-2\%$ in case of QCDF with default values for $\varphi_{H,A}^{M_i}$ and $\varrho_{H,A}^{M_i}$. 
The results are quite different between  these  approaches because of the  strong phase  mentioned
previously.

On the same figure, the asymmetry violating parameter, $a_{CP}$, is shown for the decay
$\bar{B}^0 \to \pi^+ \pi^- \pi^0$. In the vicinity of the $\omega$ resonance, the QCDF
approach gives an asymmetry of the order $-8\%$. We obtain $-20\%$ and $+5\%$ in the case of NF and QCDF
with the default values for   $\varphi_{H,A}^{M_i}$ and $\varrho_{H,A}^{M_i}$.
It appears as well that the asymmetry depends strongly on the CKM matrix parameters $\rho$ and $\eta$,
as expected.  When QCDF is applied, the asymmetry for the decay $B^- \to \pi^+ \pi^- \pi^-$, varies from 
$12\%$ to $5\%$ outside the region of the $\omega$ resonance whereas for the decay $\bar{B}^0 \to \pi^+
\pi^- \pi^0$, the asymmetry varies from $10\%$ down to $-20\%$, depending on the CKM matrix element 
parameters, $\rho$ and $\eta$. In the vicinity of the $\omega$ resonance, the asymmetry,
$a_{CP}$, takes values from $-2\%$ to $5\%$ for $B^- \to \pi^+ \pi^- \pi^-$ and from $5\%$ to 
$-30\%$ for $\bar{B}^0 \to \pi^+ \pi^- \pi^0$ when $\rho$ and $\eta$ vary. For the decay $B^- \to \pi^+
\pi^- K^-$, the asymmetry, $a_{CP}$, in the vicinity of the $\omega$ resonance, is about $+60\%$ with 
QCDF, $-40\%$ with NF and $-45\%$ with QCDF and default values for $\varphi_{H,A}^{M_i}$ and
$\varrho_{H,A}^{M_i}$. For the decay $\bar{B}^0 \to \pi^+ \pi^- \bar{K}^0$, when $\sqrt{S}$ is near 
the $\omega$ resonance, the asymmetry, $a_{CP}$ is about $+70\%$ with QCDF, $-60\%$ with
NF and $-15\%$ with QCDF and usual default values for  $\varphi_{H,A}^{M_i}$ and $\varrho_{H,A}^{M_i}$.
There is no agreement, for the value of the asymmetry between the naive  and QCD factorization at the
$\omega$ resonance except that, in both cases,  the $CP$ violating asymmetry, $a_{CP}$ reaches its maximum 
in the vicinity of $\omega$. Similar conclusions can be drawn to that of previous case  regarding
the sensitivity of the asymmetry parameter, $a_{CP}$, as well as  the CKM matrix element
parameters, $\rho$ and $\eta$.

We included $\rho^0-\omega$ mixing in order to investigate its effect on this $CP$ violating asymmetry.
The mixing through isospin violation between $\omega$ and $\rho^0$,
allows us to obtain a difference of the strong phase reaching its maximum at the $\omega$ resonance.
$\rho^0-\omega$ mixing provides an opportunity to remove  the phase uncertainty mod$(\pi)$ in the determination 
of two  CKM angles, $\alpha$ in the case of $B \to \rho^0 \pi$ and $\gamma$ in the case of $B \to \rho^0 K$.
This phase uncertainty usually arises from the conventional determination of $\sin 2 \alpha$ or 
$\sin 2 \gamma$ in indirect $CP$ violation. In QCDF, the strong phase can be generated dynamically, however,
the mechanism suffers from end-point singularities which are not well controlled. It is now apparent 
that the Cabibbo-Kobayashi-Maskawa matrix is the dominant source of $CP$ violation in flavour
changing processes in $B$ decays. The corrections to this dominant source coming from beyond 
the Standard Model are not expected to be large. In fact, the main  remaining uncertainty is
to deal with the procedure of factorization. The QCDF gives us an explicit picture of factorization 
in the heavy quark limit. It takes into account all the leading contributions as well as subleading
corrections to the naive factorization. The soft collinear effective theory (SCET) has been proposed 
as a new procedure for factorization. In the last case, it allows one to formulate a collinear
factorization theorem in terms of effective operators where new effective degrees of freedom are
involved, in order to take into account the collinear, soft and ultrasoft quarks and gluons. All 
of these investigations allow us to increase our knowledge of $B$ physics and to look for new
physics beyond the Standard Model.

\section*{Acknowledgements}
This work was supported in part by DOE contract DE-AC05-84ER40150,
under which SURA operates Jefferson Lab and by the Special Grants for
``Jing Shi Scholar'' of Beijing Normal University. 


\end{document}